\documentclass[12pt,dvips]{article}
\usepackage{epsfig}
\usepackage{hyperref}

\begin{document}

\title{``Quantum machine" to solve quantum ``measurement problem"?}
\author{Johan Hansson\footnote{\href{mailto:c.johan.hansson@ltu.se}{c.johan.hansson@ltu.se}} \\
 \textit{Division of Physics} \\ \textit{Lule\aa University of Technology}
 \\ \textit{SE-971 87 Lule\aa, Sweden}}

\date{}

\maketitle

\begin{abstract}
Recently a study of the first superposed mechanical quantum object (``machine") visible
to the naked eye was published \cite{Qmachine}. However, as we show, it turns out that \textit{if}
the object would actually be observed, \textit{i.e.} would interact with an optical photon, the quantum behavior should vanish. 
This, the actual observation, has long been suspected in many interpretations of quantum mechanics to be what makes the transition 
quantum $\rightarrow$ classical, but so far it has not been available for direct experimental study in a mechanical system. 
We show how \textit{any} interaction, even a purely quantum one, of sufficient strength can constitute a physical ``measurement"
- essentially the emergence of an effectively classical object - active observation thus being a sufficient but not necessary criterion. 
So it seems we have in this case of the ``quantum machine" a unique possibility to study, 
and possibly solve, the long-standing ``measurement problem" of quantum mechanics.
\end{abstract}
%\newpage
%\section{Introduction}
Recently the first macroscopic resonator, visible to the naked eye, obeying quantum mechanics was built and studied \cite{Qmachine}. 
To achieve its ground state required that the temperature $T \ll h f/k_B$, where $f$ 
is the mechanical mode frequency of the oscillator. For $k_B T \gg hf$ a classical behavior was observed. 
However, as $k_B T$ is just an energy, \textit{any} $E \gg hf$ should effectuate the transition, as explained below.
Thus, the transition need \textit{not} be the result of decoherence in the traditional sense (``temperature of environment"), 
as one fundamental quantum, \textit{e.g.} a photon, should suffice.

A way to directly test this would be to introduce a transparent window/shutter into the apparatus while still 
keeping $T \sim 0.1$ K \cite{Qmachine}. When open \textit{no} quantum superposition should be registered, as even a single 
interaction with an optical photon would destroy it.

To test the exact transition point and its characteristics one could use tunable electromagnetic radiation with frequency 
in the neighborhood of the fundamental mode of the mechanical oscillator, 6 GHz, or even a modification of the quantum electrical 
circuit employed in the original experiment \cite{Qmachine}.

Alternatively, one could create an oscillator with much greater stiffness, where the fundamental mode coincides with the 
visible spectrum ($\sim 10^6$ GHz), if one wishes to actually ``see" the transition. This would also be a way to study 
a \textit{visible} system in direct analogy to, \textit{i.e.} a practical realization of, Schr\"{o}dinger's Cat \cite{Schrodinger}.

In this way one could investigate the transition quantum mechanical $\rightarrow$ classical in detail and thus hopefully 
solve the quantum mechanical ``measurement problem".

%The fundamental difference between quantum and classical is:

%Quantum - First add quantum amplitudes (quantum interference), then square the results, giving probabilities.

%Classical - First take absolute squares of amplitudes (losing interference), then add the resulting probabilities.

%So we see that the crucial step to explain is amplitudes $\rightarrow$ probabilities, something that no amount of environmental decoherence in the ordinary sense can accomplish as the environment itself is quantum mechanical and hence follows the rule ``add amplitudes" (quantum entanglement). Exactly the opposite might actually be true: temperature may be just a ``symptom" signalling that the transition quantum $\rightarrow$ classical has occurred, \textit{not} a cause for it. Temperature is, after all, a statistical concept, itself dependent on probabilities, which only arise (in this context) when quantum $\rightarrow$ classical. Schematically: when an object is not observed/measured it behaves like a wave (quantum), when it is observed/measured it behaves like a particle (classical).

In linear theories, like quantum mechanics, superpositions exist indefinitely, so it seems that either a truly novel nonlinear ``collapse" 
mechanism must be introduced to complete physics, \textit{e.g.} along the lines of the GRW-model \cite{GRW} or something similar \cite{Hansson}, or a way 
to test if the superpositions really are there all the time, merely hidden, must be found.

The most conservative scenario is that the superpositions never disappear (the wavefunction never ``collapses") but instead become unobservable for all practical purposes.
This transition should be gradual, as opposed to a more abrupt transition in collapse scenarios.

The quantum mechanical amplitude, essentially the wave function, is the sum over \textit{all} paths \cite{FeynmanHibbs}, \textit{i.e.} over all possible quantum histories
\begin{equation}
K(a,b) = \sum_{All \, paths} \phi [x(t)]
\end{equation}
where
\begin{equation}
\phi [x(t)] = const \, e^{iS[x(t)]/\hbar}
\end{equation}
and $S[x(t)]$ is the action.

When $S \gg \hbar$ only the classical path differs substantially from zero, as all other paths destructively interfere, so the classical 
``principle of least action" $\delta S = 0$, and consequently the classical equations of motion, results. The condition $S \gg \hbar$ then being
the explicit realization of Bohr's more vaguely stated ``act of amplification" needed to result in directly observable classical quantities. 
It should not matter if $S$ is due to only one single quantum or very many (\textit{e.g.} by thermal ``environment"). 
As the action is relativistically \textit{invariant}
\begin{equation}
S = \int \mathcal{L} \, d^{4}x,
\end{equation}
where $\mathcal{L}$ is the invariant lagrangian density,
this would also resolve questions like ``in what frame of reference is the measurement to be made?". Due to invariance of $S$ the result is frame-independent. 
Generally the lagrangian density is composed of the free (non-interacting) part and the interaction, $\mathcal{L} = \mathcal{L}_{free} + \mathcal{L}_{int}$. 
When the quantum system is not interacting $\mathcal{L} = \mathcal{L}_{free}
 \rightarrow S \sim \hbar$.

From the specific example of the ``quantum machine" we see that the \textit{inter}action term in Eq. (2) is just $S_{int} = k_B T/\omega = E/\omega$, 
where $\omega = 2 \pi f$, and $f$ is its fundamental vibrational mode. So, to summarize, the criteria for the, experimentally controlled, interaction energy
are $E \gg hf$ for a physical ``measurement" (\textit{i.e.} emergence of ``classicality"), $E \ll hf$ for 
a purely quantum interaction and $E \sim hf$ for the important, and still unknown, but testable, transition between them.

As this transition quantum $\rightarrow$ classical should be universal, the same explanation should apply universally to all cases, systems and 
phenomena, \textit{e.g.} the screen of Young's double-slit experiment (fundamental interaction between photon and atom in screen), 
the superfluid $\rightarrow$ normal fluid transition in liquid helium (fundamental interaction between phonon and atom in fluid) and other quantum fluids, 
like Bose-Einstein condensates etc.

That highly excited ``Rydberg atoms" behave essentially classically can be immediately understood by the same principle as the energy is much higher than their 
ground state ($E \gg hf = E_0$), entailing a very complex superposition of very many wildly fluctuating different states resulting in almost classical behavior through Eqs. (1), (2).

Superposition of great many different fundamental excitations (frequencies) makes none of them distinct, so even if quantum mechanics does not ``collapse" it 
gets hidden in the ``music" consisting of the enormous number of different frequencies we interpret as classical physics. 
%Clear interference patterns, \'{a} la double-slit, disappear if too many distinct frequencies are present. 
At the same time resonant frequencies, $f$, 
scale inversely with the size of the system \cite{Qmachine}, explaining why truly macroscopic objects (including cats and human ``observers") 
are always perceived as classical. Moreover, the manifest quantum mechanical behavior 
will be hidden whenever $S_{int} \gg \hbar$, which can be taken to \textit{define} ``classical" objects, where the interaction contribution dwarfs $\hbar$.
A world built of essentially free quantum mechanical objects, where $S_{int} \ll S_{free} \sim \hbar$, would be forever and globally quantum (like unobserved photons 
in the double-slit apparatus, \textit{before} they hit the screen). 
But then of course no objects, or humans to observe them, would exist.
The ``wave-like" and ``particle-like" qualities of nature are also immediately understandable, ``unobserved" (\textit{i.e.} un\textit{disturbed}) objects are always quantum/wave-like
while ``observed" (\textit{i.e.} disturbed) objects are classical/particle-like.

If this gradual scenario described is true - and it should be testable using a variation of the ``quantum machine" - nothing ever collapses, quantum mechanics is truly universal, 
but as $\hbar$ is so tiny quantum mechanics and its characteristic superpositions are, for all practical purposes, usually invisible at our scale. 
Niels Bohr was fond of saying that the quantum world does not exist. Instead it is the classical world that does not really exist, as quantum mechanics always applies. 
Sometimes it's just very well hidden.


\begin{thebibliography}{45}

\bibitem{Qmachine} A.D.~O'Connell {\it et al.},
\textit{Quantum ground state and single-phonon control of a mechanical resonator},
Nature {\bf 464}, 697 (2010).
doi:10.1038/nature08967

\bibitem{Schrodinger} E.~Schr\"{o}dinger,
Naturwiss. {\bf 23}, pp.807-812; 823-828; 844-849 (1935).
English translation, \textit{The present situation in quantum mechanics},
Proc. Am. Phil. Soc., {\bf 124}, 323-38.
Included as Section I.11 of Quantum Theory and Measurement (J.A. Wheeler and W.H. Zurek, eds., Princeton University Press 1983).

\bibitem{GRW} G.C.~Ghirardi, A.~Rimini, T.~Weber,
\textit{Unified dynamics for microscopic and macroscopic systems},
Phys. Rev. D {\bf 34}, 470 (1986).

\bibitem{Hansson} J.~Hansson,
\textit{Nonlinear gauge interactions: a possible solution to the ``measurement problem" in quantum mechanics},
Phys. Essays {\bf 23}, 237 (2010).
doi:10.4006/1.3354830
arXiv:1001.3057 [quant-ph]

\bibitem{FeynmanHibbs} R.P.~Feynman, A.R.~Hibbs,
{\it Quantum Mechanics and Path Integrals} (McGraw-Hill 1965).


\end{thebibliography}
\end{document}